\documentclass[conference]{IEEEtran}
\IEEEoverridecommandlockouts

\usepackage{cite}
\usepackage{amsmath,amssymb,amsfonts}
\usepackage{graphicx}
\usepackage{textcomp}
\usepackage{xcolor}
\usepackage{booktabs}
\usepackage{algorithm}
\usepackage{amsmath}
\usepackage{algpseudocode}
\usepackage{float}
\usepackage{newfloat}
\usepackage{subcaption}
\usepackage[margin=1in]{geometry}
\usepackage{url}
\usepackage{multirow}
\usepackage{array} 

\def\BibTeX{{\rm B\kern-.05em{\sc i\kern-.025em b}\kern-.08em
    T\kern-.1667em\lower.7ex\hbox{E}\kern-.125emX}}
\begin{document}

\title{Enabling Network Policy Enforcement in Service Meshes\\
\thanks{This work was supported by NSF CNS Award 2213672.}
}

\author{
    \IEEEauthorblockN{Behrooz Farkiani, Fan Liu, Patrick Crowley}\\
    \IEEEauthorblockA{
        Washington University in St. Louis, \\ 1 Brookings Dr., St. Louis, MO, 63130, USA\\
        Emails: \{b.farkiani, fan.liu, pcrowley\}@wustl.edu
    }
}

\maketitle

\begin{abstract}
Portable service mesh implementations enable Layer 4 to Layer 7 policy enforcement across heterogeneous infrastructures, yet they depend on the underlying network's connectivity and policies. Layer 3 network policies govern IP traffic regardless of whether upper layers authorize the flow. While these policies are integral to security, correct enforcement often requires coordination across multiple teams, and achieving consistent policy behavior across heterogeneous environments is challenging. Studies show that most Kubernetes clusters do not enforce any network policies. We propose integrating Layer 3 network policy enforcement with service meshes to protect data-plane traffic in a portable, infrastructure-agnostic manner. This integration allows developers to define Layer 3-7 policies and to ensure enforcement across any infrastructure. Our solution builds an overlay Layer 3 network and enforces Layer 3 policies by routing traffic through specific policy enforcement points and applying default-deny principles with authorization keys. We prototyped our approach using Kubernetes and Istio and found that it adds less than 1ms of latency while supporting complex policies comparable to native Kubernetes network policies.

\end{abstract}


\section{Introduction}
Microservice architectures improve development speed and scalability over monolithic architectures by consolidating related modules into independently deployable services with high cohesion inward and loose coupling outward. Although microservices reduce software complexity and improve agility, converting function calls to distributed RPC and API calls increases operational complexity; developers must handle traffic routing, debugging, authorization, access control, and inter-service communication security \cite{fritzsch_monolith_2019,li_service_2019}. 

Service mesh architecture has emerged to address this complexity. Service mesh implementations handle Layer 4 to Layer 7 (L4-L7) communication logic, traffic routing, authorization policies, and observability without requiring service implementation changes \cite{li_service_2019,farkiani_service_2024}. Vanilla service mesh designs use proxies at both communication endpoints for traffic routing and policy enforcement. However, this approach incurs significant performance overhead, prompting various architectural improvements (e.g., Istio Ambient \cite{maturing-ambient}, Canal Mesh \cite{song-canal-2024}, Cilium \cite{cilium-mesh}) and efforts to provide developers with better control over data-plane components \cite{zhu_application_2023, zhu_high-level_2025, zhu_rethinking_2025, saxena_copper_2025}.

Despite recent advances, service meshes remain constrained by Layer 3 connectivity and the network policies of the underlying infrastructure. Network policies specify L3 and L4 ingress and egress rules between containers, virtual machines, and other endpoints \cite{li_kano_2023}, and they determine which IP flows are permitted regardless of higher-layer controls. Service meshes can express application-dependent authorization at L4-L7, typically for TCP traffic \cite{istio_protocol, linkerd_protocol, istio_l4, linkerd_policy}. In practice, however, deriving and maintaining these application-dependent policies is labor-intensive, error-prone, and often impractical at scale due to the number of services and their rate of change \cite{li_automatic_2021}. Consequently, when L4-L7 policies are absent, incomplete, or impractical to maintain at scale, L3 network policies provide a simpler, application-agnostic safety net by ensuring that only traffic from authorized sources can reach an endpoint.

We focus on containerized environments and Kubernetes, which holds 83\% of the market share \cite{li_automatic_2021, kubernetes_report}, though our approach applies to other settings. Kubernetes clusters default to unrestricted pod-to-pod communication, and network policies enable network segmentation by blocking unauthorized traffic \cite{brendel_high_2016}. These policies should be enforced by network plugins (i.e., CNI plugins \cite{kubernetes_plugin}), though some ignore them (e.g., \cite{flannel}). Despite being critical for service security, and despite extensive research on automation and verification \cite{kim_kubeaegis_2024, pizzato_intent-based_2024, li_kano_2023}, various challenges prevent wide-scale adoption of network policies; policy handling varies dramatically across network plugins \cite{kim_exploring_2025}, and consistent enforcement across heterogeneous and multi-cloud infrastructures is particularly challenging \cite{kim_kubeaegis_2024, kim_exploring_2025}. Beyond technical complexity, organizational challenges compound the problem: developers and infrastructure teams have different objectives and must coordinate closely to implement policies correctly. These combined challenges result in remarkably low adoption, with only 9\% of Kubernetes clusters implementing network policies \cite{pizzato_intent-based_2024, wiz}.

This work is, to our knowledge, the first to present data-plane components that let developers enforce Layer 3 policies in a portable, infrastructure-agnostic way entirely within service meshes, thus addressing the above challenges. We integrate with portable service meshes that capture traffic and enforce policies inside each workload's network namespace rather than relying on host-level redirection (e.g., Istio \cite{sidecar_ambient, maturing-ambient} and Linkerd \cite{linkerd_arch}). These meshes require only basic connectivity support from infrastructure, which is provided by almost all infrastructures. Integration with our approach enables them to apply L3-L7 policies across diverse environments because our approach requires no infrastructure-level support. In addition, unmanaged endpoints can also be secured. For example, an end-user device can join the extended Layer 3 overlay, enabling end-to-end Layer 3-7 traffic management and security.

Our approach creates tunnel interfaces for each overlay entity, forming an L3 overlay that carries service traffic and enables data-plane network policy enforcement. Traffic between overlay entities passes through Stand-alone Proxies (\textbf{SaP}s), which enforce network policies by matching overlay IPs against an Access Control List (ACL). This enables L3 policy enforcement outside entity namespaces, similar to infrastructure-level policy enforcement. Additionally, we employ a default-deny policy combined with authorization keys. Entity pairs must be explicitly allowed to communicate, with the control plane distributing corresponding authorization keys. Our approach also installs iptables rules that allow incoming traffic only from overlay interfaces. This prevents unauthorized traffic from reaching service endpoints via infrastructure and ensures all incoming traffic carries authorization keys and passes through SaPs. We prototyped our solution using Kubernetes \cite{kubernetes} and Istio \cite{istio} as the infrastructure and service mesh. We demonstrate that our portable network policy enforcement can enforce complex L3 policies. We also show that our approach adds less than 1ms of overhead and is transparent to service-level L4-L7 logic and policies, allowing service components to use strict mTLS and maintain secure end-to-end connections.

The paper is organized as follows: Section \ref{question} describes the solution components and how it enables L3 policy enforcement. Section \ref{imp} presents the evaluation, and Section \ref{conclusion} concludes the paper.

\section{Solution and Implementation}\label{question}
We add two data-plane components to enable L3 network policy enforcement, both controlled by the service mesh control plane.

\textbf{The L3 component:}
A container running inside overlay-connected entities (e.g., pods, VMs, end-user devices) and configured by the control plane. It performs the following functions: 1) creates at least one tunnel device and assigns IP addresses determined by the control plane, 2) captures overlay-related L3 traffic, 3) reads from and writes to the tunnel, 4) encapsulates/decapsulates overlay L3 traffic in GUE \cite{GUE} over UDP using a designated UDP port for communication, 5) appends a 32-bit authorization key to the GUE header for outgoing traffic and verifies keys for incoming traffic, 6) uses a forwarding table populated by the control plane. The encapsulation overhead is 8 bytes for GUE plus UDP and IP headers, totaling 36 bytes for IPv4. End-users outside the Kubernetes cluster can either run the L3 component on their devices to join the extended L3 overlay or interact with overlay entities through SaPs that are acting as ingress/egress gateways.

\textbf{SaPs:}
SaPs are pods with an L3 component, an ACL, and a forwarding table. SaPs enforce policies by checking packets against the ACL, which is populated by the control plane, to verify that overlay source and destination IPs are authorized to communicate in each direction. They then forward traffic based on the overlay topology. If the next hop and the SaP are both part of the overlay, the SaP uses UDP forwarding. If either the source or destination is not an overlay entity (i.e., does not have an L3 component), the SaP couples with a proxy and acts as an ingress or egress gateway. SaPs can forward traffic to other SaPs, allowing multiple SaPs between the source and destination.

The overlay control plane, as part of the service mesh control plane, creates a flat IPv4 or IPv6 overlay network in each Kubernetes cluster by assigning unique IP addresses to L3 component tunnel interfaces. The address range assigned to overlay entities is configured not to conflict with IPs assigned to entities by the infrastructure. For example, the overlay can use IPv6 ULA \cite{haberman_unique_2005} addresses or IPv4 carrier-grade NAT addresses (as used in Tailscale \cite{tailscale}). An overlay network can span multiple Kubernetes clusters if its address range does not conflict with cluster addresses, and communication between overlays is through SaPs. The control plane assigns an authorization key to each overlay entity. When communicating with a destination, this authorization key must be appended to the GUE header as part of the group identifier field. We selected GUE for its lightweight, connectionless design that does not interfere with underlying IP logic \cite{herbert_extensions_2019}.

Since IP addresses assigned to tunnel interfaces exist only inside entities and are invisible to infrastructure, both the L3 component and SaPs use a forwarding table that maps overlay IPs to a next hop with an infrastructure-assigned IP. The forwarding table also contains authorization keys. By default, the next hop for each source-destination pair is set to the infrastructure-assigned IP of an SaP instance to ensure traffic passes through SaPs. Although the next hop points to an SaP, the authorization keys in the forwarding table belong to the intended destination, not the SaP. Entities are unaware of SaPs, making overlay updates easier. For example, for performance reasons, the next hop can be updated to point directly to the intended destination, enabling direct communication between overlay entities without passing through SaPs, though this requires explicit developer configuration. Consequently, SaPs do not modify authorization keys; they only verify that overlay entities are allowed to communicate based on ACL rules, then forward data to the next hop according to their forwarding table, which may be the destination overlay IP or another SaP. The ACL rules contain overlay IPs and any other information necessary for policy enforcement.

The service traffic needs to be automatically sent to the overlay without requiring modification of services. This can be done by mapping endpoint names (like service names) to overlay IPs. For endpoints that are discoverable through the service mesh, we need to modify the discovery mechanism to return overlay IPs instead of infrastructure IPs. Another approach is to use a proxy inside the entity namespace that can capture DNS requests \cite{dns_proxy} and return overlay IP addresses. Alternatively, Kubernetes headless services \cite{kubernetes_service} can globally map service names to overlay IPs. All these approaches remain transparent to L4-L7 service logic. Through these approaches, endpoints automatically send traffic to overlay IPs, which are then captured and processed by the tunnel interface transparently.

To limit data-plane traffic to authorized flows only, we employ a default-deny model: endpoints may not communicate until a policy explicitly authorizes communication. When a policy is created, the overlay control plane updates the forwarding tables of authorized endpoints by adding corresponding authorization keys and next-hop addresses. Additionally, when an entity joins the overlay, the L3 component installs iptables rules that drop data-plane traffic destined for service ports if it arrives on non-overlay interfaces.

The overlay control plane is part of the service mesh control plane, configuring tunnel parameters and forwarding tables for overlay entities. Although authorization keys need not be unique across the system, the control plane rotates them regularly. Additionally, it configures the ACL and forwarding logic for each SaP and monitors overlay performance to determine next-hop addresses. The control plane must populate the forwarding table in a way that minimizes latency incurred by policy enforcement.

Although SaPs only forward overlay data without modification, passing all traffic through SaP instances introduces latency overhead depending on SaP location relative to traffic sources. The control plane should control and monitor SaP locations and overlay latency. A simple deployment strategy runs one SaP pod per node using a DaemonSet \cite{daemonset}, where each instance enforces policies for local node traffic. This reduces latency by keeping enforcement close to workloads. SaPs can be co-located with the mesh's on-node data-path component, for example the Istio CNI DaemonSet \cite{istio_cni}.

To better understand how the overlay prevents unauthorized access, we consider a spoofing attack with extensive information access. An unauthorized entity obtains another entity's valid overlay IP, infrastructure-assigned pod IP, and authorization keys, then sends traffic encapsulated in GUE directly to the destination while bypassing SaPs. We assume the unauthorized entity can send packets using an overlay IP already authorized to communicate with the destination, since the destination checks its forwarding table to respond and it only contains authorized overlay IPs. We also assume the application layer is not vulnerable, as network policies do not concern application logic. Thus, the destination processes the traffic and returns the response to an SaP. When the SaP receives return traffic, it checks the packet against its ACL. Even if the packet passes ACL checks, the SaP forwards it to the entity that authentically owns the spoofed overlay IP based on its control-plane-populated forwarding table, not the unauthorized sender. While this attack requires extensive information and an ill-configured ACL, return traffic will not reach the unauthorized sender. Additionally, regularly rotating authorization keys prevents this attack.

\begin{figure}[htbp]
  \centering
  \includegraphics[width=0.8\columnwidth]{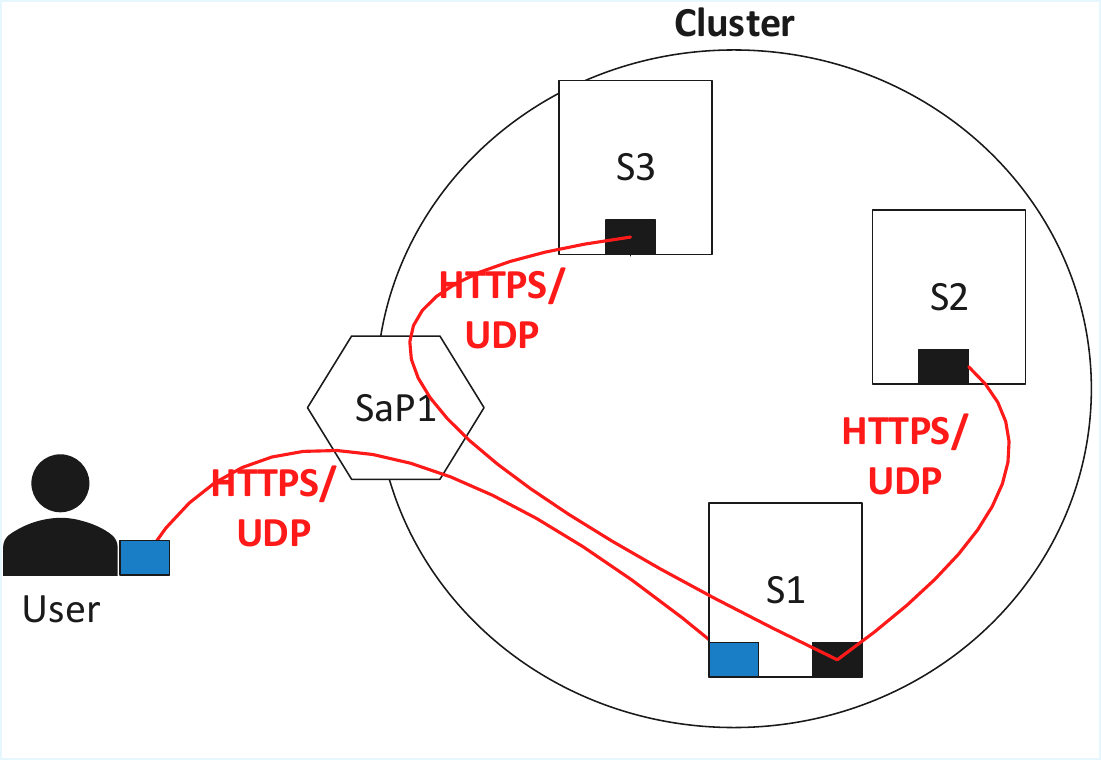}
\caption{Overlay evaluation topology consisted of four pods. There are two overlay subnets (blue and black) and the tunnel devices that are part of each one are shown with the same color.}
  \label{fig:example}
\end{figure}

\section{Prototype Implementation and Evaluation}\label{imp}
To evaluate the feasibility and performance of our approach, we implemented a prototype using Kubernetes v1.30.13 with Calico v3.28.1 \cite{calico} and Istio v1.26.0 in sidecar mode as the service mesh, and we integrated our solution with Istio. Since Calico can enforce Kubernetes network policies, we compare the performance of our approach with a native Kubernetes approach. 

\begin{table}[htbp]
\centering
\caption{L3 network policies and next hops}
\label{tab:routing-table}
\begin{tabular}{|l|l|l|}
\hline
\multicolumn{1}{|c|}{\textbf{Endpoint}} & \multicolumn{1}{c|}{\textbf{Destination}} & \multicolumn{1}{c|}{\textbf{Next Hop}} \\ \hline
\multirow{3}{*}{S1} & S2 & S2 \\ \cline{2-3}
& S3 & SaP/Nginx \\ \cline{2-3}
& End-User & SaP/Gateway \\ \hline
S2 & S1 & S1 \\ \hline
S3 & S1 & SaP/Nginx \\ \hline
End-user & S1 & SaP/Gateway \\ \hline
\end{tabular}
\end{table}

\begin{table}[htbp]
\centering
\caption{Average HTTP latency(ms) results. Standard deviations are shown in parentheses.}
\label{tab:latency-results}
\begin{tabular}{l|l|l|l|}
\cline{2-4}
& \textbf{S1-S2} & \textbf{S1-S3} & \textbf{User-S1} \\ \hline
\multicolumn{1}{|l|}{Our Approach} & 1.8(0.3) & 2.1(0.5) & 4.5(0.5) \\ \hline
\multicolumn{1}{|l|}{Native} & 1.2(0.2) & 5.1(0.8) & 7.7(0.7) \\ \hline
\end{tabular}
\end{table}

We implemented three simple HTTP echo web servers. Each server issues an HTTP/1.1 request to a configured peer. SaP and the L3 component are Python modules configured during pod startup by an init container, which enforces the correct configuration for each pod based on its role. The init container configures the L3 component, writes forwarding tables per Table \ref{tab:routing-table}, programs the SaP ACL, and adds iptables rules that drop non-overlay traffic. The L3 component also creates a character device using mknod \cite{mknod}, then creates and configures tunnel interfaces, requiring NET\_ADMIN and MKNOD capabilities without privileged access.

We used one cluster with one SaP, three services (S1-S3) with one instance each, and one external user, as shown in Figure~\ref{fig:example}. The end-user ran Envoy proxy v1.33.3 \cite{envoy} and the L3 component, communicating with S1 through SaP. We configured SaP1 as a load balancer that acts as a UDP forwarder to enable S1 $\leftrightarrow$ S3 and end-user $\leftrightarrow$ S1 communication without using its L3 component. S1 has two tunnel interfaces since it participates in two overlays: one with IPv6 for communication with end-users, and another with IPv4 within the same subnet as other services. Only the communication paths listed in Table~\ref{tab:routing-table} were allowed; for example, S2 could not communicate with S3. Each service S1-S3 was equipped with an Istio proxy. However, since the Istio mTLS configuration could not be applied to the end-user proxy, we added an Envoy proxy container inside the S1 pod to enable mTLS communication between the end-user and S1. Because we had not yet implemented a service mesh control plane component, we used Kubernetes headless services \cite{kubernetes_service} for services S1-S3. These services did not have cluster IPs, and for each service, we defined an endpoint with the overlay IP, allowing Istio to discover only overlay IPs. We set PeerAuthentication to STRICT \cite{istio_peer} and DestinationRule to ISTIO\_MUTUAL \cite{istio_destination}. Although the communication between S1$\leftrightarrow$S2 does not pass through SaP1 in this setup, this approach requires explicit setup because it weakens policy enforcement by relying solely on authorization keys.

We compared our implementation with a native Kubernetes/Calico implementation. We used an Istio ingress gateway with simple TLS mode to connect the end-user to the cluster. The gateway terminated TLS and sent traffic to S1, splitting the connection, whereas in our approach TLS is terminated at S1. We also used an nginx proxy in place of our SaP to prevent direct S1-S3 communication; the proxy terminated connections and contacted the other party. We used Kubernetes network policies and only allowed communication between S1$\leftrightarrow$S2, S1$\leftrightarrow$proxy, S3$\leftrightarrow$proxy, and S1$\leftrightarrow$gateway for ingress traffic. PeerAuthentication and DestinationRule were similar to our approach.

The average latency results based on at least 30 samples are shown in Table \ref{tab:latency-results}. In both experiments, all pods were created on the same node to eliminate inter-node overhead. Comparing direct S1$\leftrightarrow$S2 and indirect S1$\leftrightarrow$S3 communication in our approach, passing traffic through SaP increased latency by 0.3ms, although this mostly depends on SaP placement. Comparing S1$\leftrightarrow$S2 and S1$\leftrightarrow$S3 with the native S1$\leftrightarrow$S2 latency, the overlay encapsulation/decapsulation increases latency by less than 1ms. Since S1$\leftrightarrow$S3 and end-user$\leftrightarrow$S1 connections are end-to-end in our approach, we observe that our approach has 3ms lower latency than the native approach.

\section{Conclusion}\label{conclusion}
We presented a portable, infrastructure-agnostic mechanism for Layer 3 policy enforcement that integrates with service meshes to secure data-plane traffic independently of higher-layer logic. Our data-plane components enable developers to enforce Layer 3 alongside Layers 4-7 policies in a unified way across heterogeneous environments. We prototyped the design and showed that, with minimal overhead, our solution enforces diverse policies with effectiveness comparable to infrastructure-native enforcement.

This work presents a proof of concept that enables portable L3 data-plane policy enforcement in service meshes. As future work, we will design control plane components to compute optimal SaP placement. Our prototype integrates with Istio in both sidecar and ambient modes. However, we did not report ambient results here because the experimental setup required setting PeerAuthentication to PERMISSIVE to allow plaintext communication. A more mature implementation will provide robust integration with multiple service mesh implementations.

\bibliographystyle{IEEEtran}
\bibliography{refrences}
%


\end{document}